\documentclass[twocolumn,noshowpacs,preprintnumbers,amsmath,amssymb]{revtex4}

\usepackage{verbatim}

\usepackage{graphicx}
\usepackage{dcolumn}
\usepackage{bm}
\usepackage{wrapfig}

\begin{document}

\title{Quantum Dot Attachment and Morphology Control by Carbon Nanotubes}
\author{Beatriz H. Juarez}
\author{Christian Klinke}
\author{Andreas Kornowski}
\author{Horst Weller}
\email{weller@chemie.uni-hamburg.de}
\affiliation{Institute of Physical Chemistry, University of Hamburg, D - 20146 Hamburg, Germany}

\begin{abstract} 

Novel applications in nanotechnology rely on the design of tailored nano-architectures. For this purpose, carbon nanotubes and nanoparticles are intensively investigated. In this work we study the influence of non-functionalized carbon nanotubes on the synthesis of CdSe nanoparticles by means of organometallic colloidal routes. This new synthesis methodology does not only provide an effective path to attach nanoparticles non-covalently to carbon nanotubes but represents also a new way to control the shape of nanoparticles.

\end{abstract}

\maketitle

There is an increasing number of potential applications for materials with dimensions in the nanometer range. Such systems show improved or even new properties emerging as a result of electronic confinement. In particular, carbon nanotubes (CNTs) find applications as transistors~\cite{C01,C02}, conductive layers~\cite{C03}, field emitters~\cite{C04,C05},  and mechanical components~\cite{C06,C07,C08}. At the same time, the knowledge acquired from well-established synthetic procedures has facilitated the tailoring and optimization of semiconductor nanoparticles (NPs) with efficient, photostable luminescence properties controlled by quantization~\cite{C09,C10,C11}. A high degree of size and shape control has been achieved during the last decades for some specific systems such as CdSe~\cite{C09}. Furthermore there is a strong interest to attach NPs to one-dimensional systems like CNTs. For example, metallic particles may serve as catalysts to create branches on nanotubes~\cite{C12}, while semiconducting NPs can act as light absorbing sites to increase the photoconductivity of CNTs~\cite{C13}. In previous studies, semiconductor NPs have been grown on CNTs by generation of defects in the CNT lattice structure by means of covalent functionalization~\cite{C14} or ozonolysis~\cite{C15}. These aggressive treatments render an oxidized CNT surface or even a structural damage which deteriorates their outstanding electrical, mechanical, and optical properties significantly~\cite{C16,C17}. Other strategies consist in an adsorption of surface active molecules, which in turn may bind nanoparticles electrostatically~\cite{C18,C19} or the self assembly of CdSe and InP nanoparticles~\cite{C20} using the grooves in bundles of single wall carbon nanotubes (SWNTs) as a one-dimensional template.

Here, we present a novel one pot synthesis approach in which semiconductor NPs can be specifically attached to non-functionalized and non-pretreated CNTs at a very high degree of coverage, and which, in addition lead to well defined morphological transformations of the NPs. The attachment is observed for CdSe NPs on both, singlewall and multiwall carbon nanotubes (MWCNTs). Furthermore, the obtained composite materials exhibit photoelectrical response. 

\begin{figure}[!h]
\begin{center}
\includegraphics[width=0.45\textwidth]{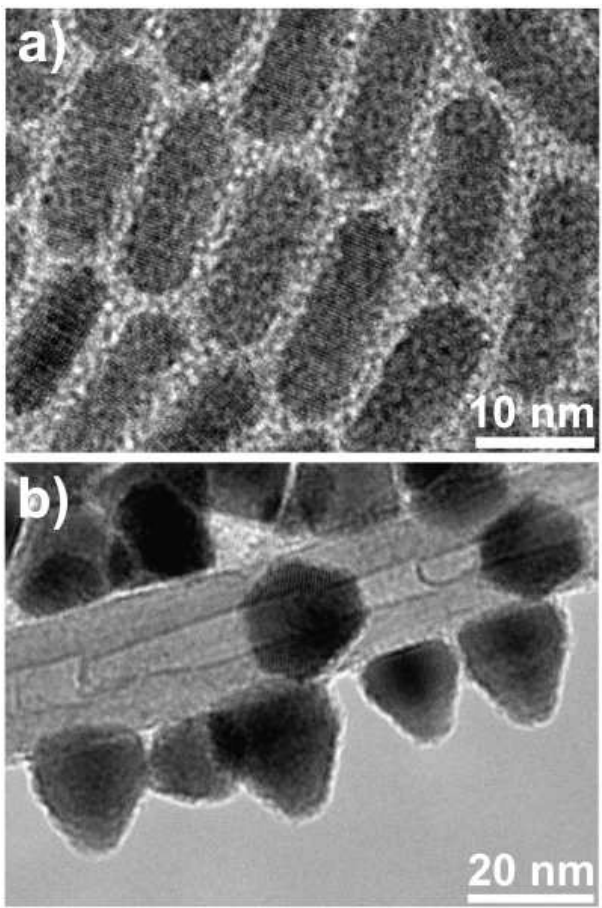}
\caption{\it CdSe particles obtained after 48~h hours (a) in absence and (b) in the presence of MWCNTs.}
\label{F01}
\end{center}
\end{figure}

CdSe nanorods were synthesized following a method similar to the one introduced by Peng et al.~\cite{C21}. In a standard procedure cadmium oxide CdO (0.025~g) was complexed by octadecylphosphonic acid ODPA (0.14~g) in tri-n-octylphosphin oxide TOPO (2.9~g) after degassing the mixture at 120$^{\circ}$C for one hour. Once an optically clear solution was obtained, pure selenium dissolved in tri-n-octylphosphine TOP (0.42~mL 1M) was injected (above 250$^{\circ}$C), promoting the nucleation of CdSe particles after a few minutes. The growth temperature reaction is 245$^{\circ}$C. CdSe nanorods were formed in a few hours. Figure~\ref{F01}a shows a transmission electron microscopy (TEM) image of CdSe NPs obtained under these conditions after 48~h. In order to study the influence of CNTs on the synthesis of NPs, non-functionalized CNTs suspended in 1,2-dichloroethane (DCE) were injected to the complexed Cd solution at 70$^{\circ}$C. The nanotubes suspension was prepared by sonication in DCE for 30~min. After the injection, the solvent was evaporated under vacuum for about 15~min prior to Se injection. In the course of the here presented studies different types of CNTs have been used: SWCNTs produced by the HiPCO method and CVD (Nanocyl), SWCNTs grown by laser ablation, as well as MWCNTs obtained by means of CVD (Baytubes). 

When the same experiment was carried out in the presence of a small amount of CNTs (0.2 - 0.6~mg), pyramidal CdSe particles were observed closely attached to the CNTs (Figure~\ref{F01}b). Very similar results were found for both SWCNT and MWCNT. The degree of NP coverage was substantially higher than that obtained by means of crude methods involving surface oxidation of the carbon lattice~\cite{C15,C16}. High resolution transmission electron microscopy (HRTEM) investigations also exhibit an atomically close distance between NPs and CNTs leaving no space for the original capping ligands. The attachment is, therefore, interpreted in terms of a direct tight binding between NPs and CNTs. The high degree of coverage excludes hereby the possibility that this binding occurs only at occasionally formed defect sites of the CNTs. 

\begin{figure}[!h]
\begin{center}
\includegraphics[width=0.45\textwidth]{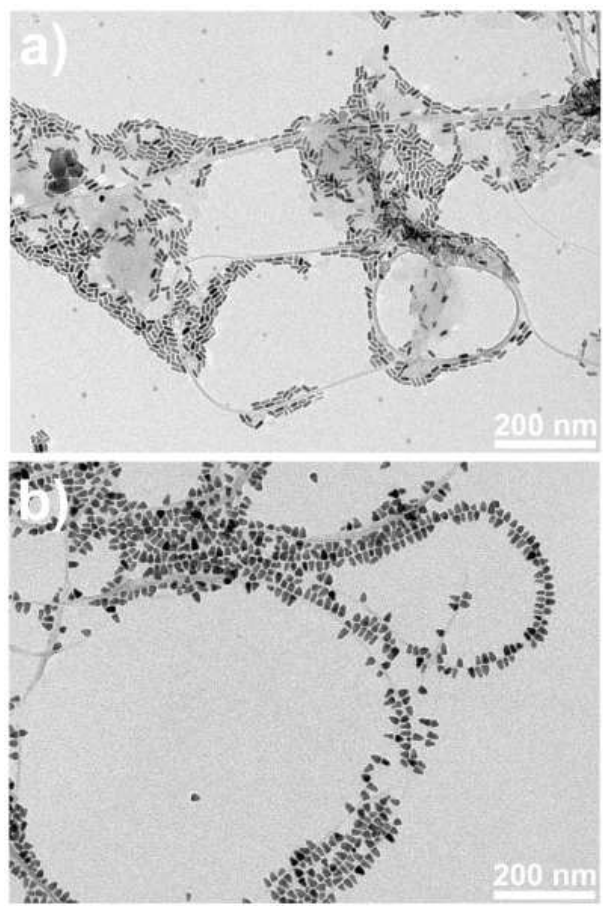}
\caption{\it Synthesized CdSe nanoparticles in the presence of 0.6~mg of SWCNT after (a) 2~h and (b) 60~h.}
\label{F02}
\end{center}
\end{figure}

We followed the evolution of the particles and found that in early stages of the growth (Figure~\ref{F02}a) no significant deviations can be observed with regard to the synthesis in absence of CNTs. The loose aggregation of nanorods around the CNTs is a typical result of drying the sample on the TEM grid and does not indicate any specific interaction between CNTs and nanorods in this early stage of genesis. Contrary, after longer periods of time depending on the type (single or multiwall) and the amount of CNT, drastic changes in shape and attachment of such particles are apparent (Figure~\ref{F02}b). The quantum rods undergo a morphology change to pyramidal particles with a size ranging from 10 to 20~nm (c-axis). Worth mentioning is that particles not attached to the CNTs possess pyramidal form as well. The CNT-NP composites were washed by several cycles of centrifugation and sonication in organic solvents. We observed that the system evolves faster in the presence of SWCNT (regardless the source) compared with MWCNT. We attribute this to the higher reactivity of SWCNTs compared to MWCNT, which is based on the stronger curvature in the graphitic structure and manifested in several examples including defect and sidewall functionalization~\cite{C15}.

Shape controlled CdSe nanoparticles including tetrapods, arrows and tearpods, is obtained by anisotropic growth of different facets influenced by the ratio of ligands, temperature and monomer concentration~\cite{C22}. According to previous studies~\cite{C23} the shape evolution of CdSe dots under the described experimental conditions (in the absence of CNT) occurs in three different stages. They grow in a diffusion-controlled regime, a scenario where monomers migrate into the diffusion sphere of the nanoparticles and are mainly consumed by the facets perpendicular to the c-axis (1D-growth). If the reaction proceeds for longer times 3D growth and 1D-2D ripening processes may lead to a more rounded shape~\cite{C24}. This feature was observed in our studies for CdSe nanorods when the reaction without nanotubes proceeds for longer times (i.e. 48~h, rounded caps) as shown in Figure~\ref{F01}a. However, pyramidal shaped particles are not observed in the absence of CNTs. 

Other studies have demonstrated an evolution from hexagonal to a pyramidal shape in an understoichometric sulphur/cadmium medium for CdS nanoparticles~\cite{C24}. Moreover, with extra injections of both Cd and Se precursors, a similar morphological transformation from rods to pyramids has recently been reported for CdSe nanoparticles~\cite{C25}.

\begin{figure}[!h]
\begin{center}
\includegraphics[width=0.45\textwidth]{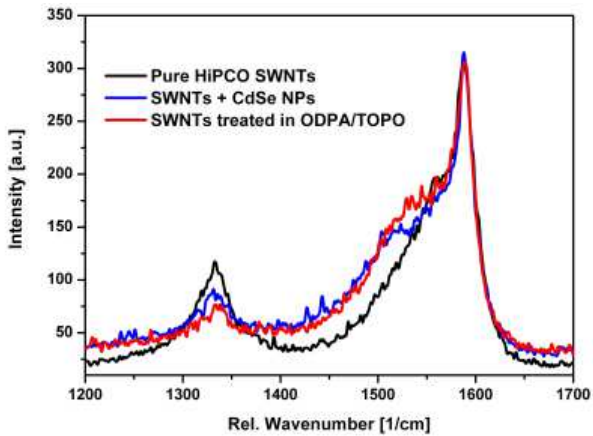}
\caption{\it Raman response of raw SWCNTs (HiPCO) before the synthesis (black), SWCNTs treated in a mixture ODPA/TOPO at 245$^{\circ}$C (red) and that of the sample shown in Figure~\ref{F02}b (blue). Raman excitation wavelength: 514.5~nm. Laser power density on the sample: 0.3~kW/cm$^{2}$.}
\label{F03}
\end{center}
\end{figure}

In order to investigate the nature of the chemical bonds between the NPs and CNTs the Raman spectra were investigated, which allowed a clear identification of sp$^{2}$ and sp$^{3}$ carbon. Figure~\ref{F03} displays the Raman response of pure SWCNTs before the synthesis and that of the sample shown in Figure~\ref{F02}b. Additionally, a spectrum of a blank sample is shown for which the nanotubes were treated in a mixture of ODPA/TOPO at 245$^{\circ}$C, but in absence of both Cd and Se sources. Most striking is the similarity of the D peak intensities around 1330~cm$^{-1}$, which are attributed to sp$^{3}$ carbon, for all three samples. The slightly larger signal of the pure CNTs is probably due to carbon contaminations which have been removed during the chemical treatment of the other samples. These findings clearly evidence that the carbon lattice is neither oxidized (sp$^{3}$ hybridizated) by the treatment with ODPA/TOPO, nor during the particle evolution and attachment. It also implies that the NPs cannot form a covalent bond, e.g. between Se and carbon, since this would also require sp$^{3}$ carbon. We, therefore, propose that particle attachment is based on a non-covalent bond formed by an interaction of the CNT pi-system with Cd-rich NP facets. In the Raman spectra of the ODPA/TOPO treated CNTs as well as for the ones with attached NPs, we furthermore observe a broadening of the double G peak around 1600~cm$^{-1}$ attributed to carbon in the sp$^{2}$ state. Similar broadening effects have been observed for reversibly charge loaded CNT films.  Thus, a charge transfer associated with OPDA or Cd-rich NP facets attached to the CNTs is conceivable. 

The finding that the sp$^{2}$ carbon structure and the thereof resulting electronic structure of the nanotubes is not significantly modified by the nanoparticle attachment is also supported by the absorption spectra in the NIR region, which exhibit no noticeable changes in the dipole active exciton transitions E11 and E22~\cite{C27,C28,C29}. A covalent attachment would lead to a drastic reduction or disappearance of the absorption bands~\cite{C18}.

\begin{figure}[!h]
\begin{center}
\includegraphics[width=0.45\textwidth]{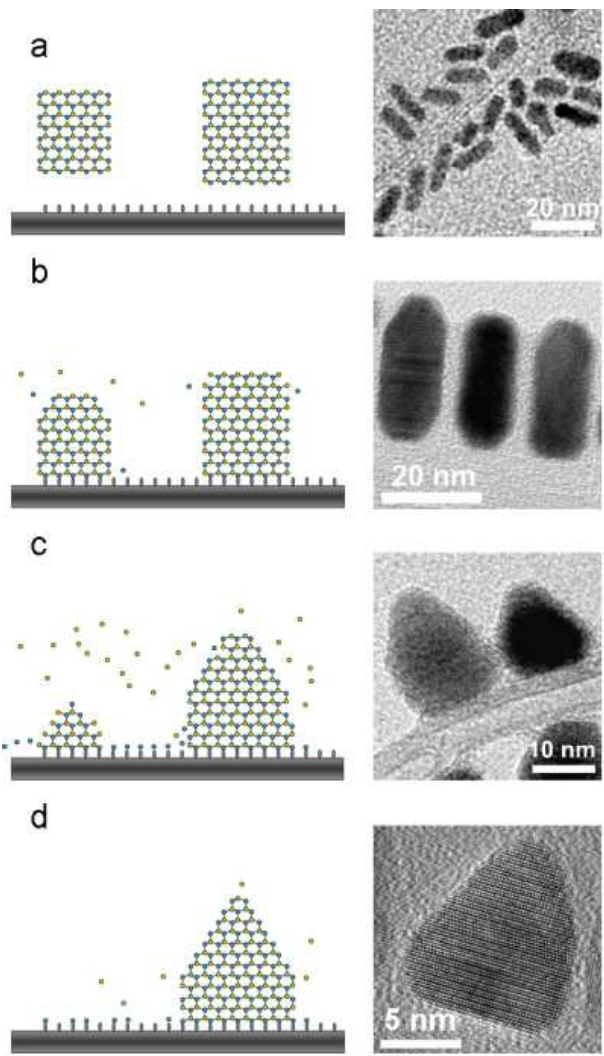}
\caption{\it Proposed mechanism for the interaction and morphological transformation of CdSe nanorods in the presence of CNTs (with corresponding TEM example images). (a) NP-CNTs interaction, (b and c) selective etching, (c and d) selective etching and Ostwald Ripening, (d) final shape of the pyramidal nanoparticles.}
\label{F04}
\end{center}
\end{figure}

Figure~\ref{F04} illustrates the proposed mechanism for the interaction of CdSe with CNTs and further morphological transformation along with corresponding TEM images. In first stages of the reaction (a) CdSe nanorods are formed and remain in solution capped with ODPA and TOP as in the absence of CNTs, whereas the CNTs possibly can be surrounded by the organic ligands. From a molecular point of view the attachment of the nanorods can be understood in terms of a ligand exchange of ODPA from the Cd sites of the nanorods against the carbon nanotubes. It implies that CNTs can be treated like coordinating ligands during nanoparticle synthesis opening novel concepts in nanochemistry. This is also supported by the fact that the attached NP can be detached from the CNTs by the addition of stronger ligands to the composite suspension, e.g. like thiols. The here described concept also explains the preferential orientation of the nanorods with their c-axis perpendicular to the surface of nanotubes, since in this orientation the Cd-rich $\{$001$\}$ lattice planes may be exposed to the pi-system of the nanotubes. Whether, however, the interaction is purely electrostatic or based on the overlap of the delocalized $\pi$-orbitals of the CNTs with the empty conduction band levels of the NPs formed by Cd cannot be distinguished at the present state of experiments, but is probably an interesting problem for theory. Nevertheless, binding of NPs at the CNTs must be rather strong, since treatment in an ultrasonic bath is not sufficient to remove the particles from the CNT surface. 

\begin{figure}[!h]
\begin{center}
\includegraphics[width=0.45\textwidth]{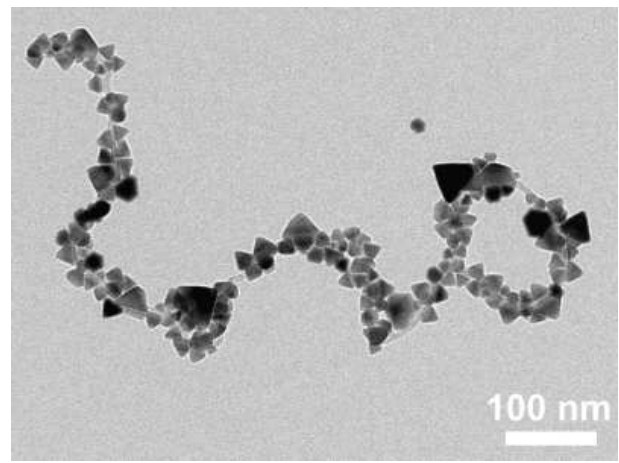}
\caption{\it SWCNT-CdSe NP composites treated for 1~week. A wider size distribution of particles attached to the carbon lattice is apparent due to the Ostwald ripening mechanism.}
\label{F05}
\end{center}
\end{figure}

The shape transformation (b) starts with an etching process of ODPA, which was also proven in blank experiments without CNTs. The ODPA etching should lead primarily to a Cd release starting at the edges of the nanorod (either at the top or at the bottom). However, at the bottom the charge density should be increased by the bond to the nanotubes and, as a consequence, Cd may be stabilized towards the etching. Of course the particles will also slowly release Se in order to retain the proper stoichiometry (c). The preferential Cd etching should gradually convert all the etched planes into Se-rich ones terminated by TOP. In a stoichiometric crystal the $\{$10-1$\}$ family of planes consist of either Cd or Se, respectively. However, Se-rich surfaces should be favored under dynamic etching conditions, driving the particles into the finally observed morphology. The formation of these planes is in agreement with the angle of approximately 30$^{\circ}$ measured in the TEM images between the $\{$10-1$\}$ planes and the c-axis of the formed pyramids. In the wurtzite structure this angle is 28$^{\circ}$. This model is in accord with ongoing tomographic TEM studies~\cite{C30}. Since the CNTs have a strong affinity to Cd the released Cd-ODPA species should be enriched and able to easily diffuse at the surface of the nanotubes to the neighboring particles. This diffusion of Cd monomers on the surface of the nanotubes and the growing of facets with low chemical potential (big particles) lead to the dissolution of some particles and the growing of some others, following the classic concept of Ostwald ripening (c,d)~\cite{C31}. This explanation is consistent with samples treated for 1 week under the mentioned experimental conditions (Figure~\ref{F05}) in which a wider size distribution of particles attached to the carbon lattice is apparent. The dramatic shape transformations after one week of Ostwald ripening implies that some lateral mobility of the atomic constituents of the nanoparticles at the CNT's surface must be possible. Furthermore, the fact that pyramidal-shaped particles were also found not only in contact to the nanotubes but also in solution suggests a mechanism of attachment and de-attachment from the carbon lattice. As determined by HRTEM and selected area electron diffraction (SAED) patterns, the crystal lattice in the pyramidal CdSe NPs remains hexagonal after the transformation.

These results suggest that other carbon allotrope surfaces can also have a clear influence. Preliminary experiments with C60 fullerenes and graphite yield indeed very similar shape transformations of the initially formed CdSe nanorods into pyramidal configurations and attachment to the carbon surfaces. Furthermore, similar shape transformations take place in the presence of non pi-systems like SiO$_{2}$ microspheres. The influence of SiO$_{2}$ microspheres in the CdSe nanorod synthesis yields a similar NP transformation but in contrast to the pi-systems without attachment. Thus, on the one hand, the shape transformation is not limited to CNTs but to systems capable to modify the coordination sphere of the NP. This statement is supported by other work where the replacement of TDPA ligands for more labile ones (acetate) promotes a shape transformation from rods to pyramidal-base particles~\cite{C26}. On the other hand, the stable attachment to obtain a composite material requires the $\pi$-orbitals of systems like CNTs or graphite. 

\begin{figure}[!h]
\begin{center}
\includegraphics[width=0.45\textwidth]{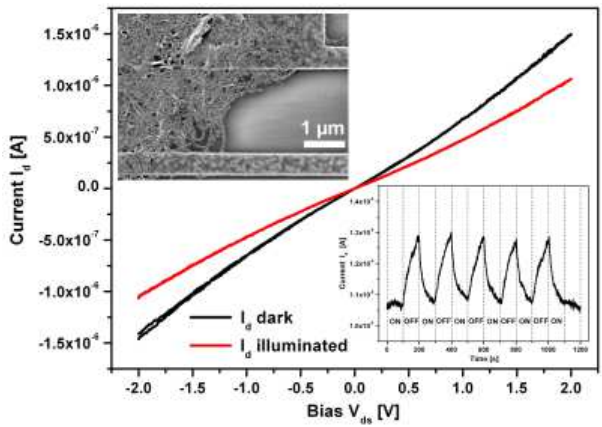}
\caption{\it Typical I$_{d}$-V$_{ds}$ characteristic of the CNT-CdSe composite bridging two gold electrodes on SiO$_{2}$. Upper inset: wavelength-resolved current of the device structure. Lower inset: Development of the current with time under chopped illumination without filtering.}
\label{F06}
\end{center}
\end{figure}

We also investigated the electrical response of the CNT-CdSe composite system. A typical current-voltage characteristic with and without illumination is shown in Figure~\ref{F06}. The lower inset shows a time profile of the drain current under light chopping (150 W cold light halogen lamp Schott KL 1500-Z) at a bias of V$_{ds}$ = 2 V whereas the  wavelength-resolved current is depicted in the upper inset. The most remarkable feature is the current decrease under illumination. This decrease lies above 700~nm, in agreement with the wavelength of the absorption edge of CdSe. These finding indicates a strong electronic coupling between the NPs and the CNTs, as it is expected from the tight attachment. Furthermore, the curves show practically no hysteresis and a slight nonlinearity due to possible tunnel barriers at the contacts. At the present state of investigations it is difficult to interpret the photoelectrical response, since the conduction band level of CdSe is very similar to the work function of the nanotubes. However, the observed behavior can be explained in terms of a gate effect and/or an enhanced electron-hole recombination within the CNT. During illumination excitions are generated in the NPs. Electron transfer from the NP to the conducting CNTs might then occur resulting in an enhanced electron-hole recombination within the nanotubes and thus a reduction of the current from contact to contact. The holes in the NP would remain in the nanoparticles and slowly be withdrawn via reactions with the environment. The remaining positive charges in the nanoparticles could furthermore have an electrostatic influence on the semiconducting SWNTs, a gate effect. Positive charges can drive semiconducing SWNT into the off-state and the current flow is drastically reduced~\cite{C18}. It will be very interesting to investigate the electric transport properties of CNT-NP composites in more detail. 

In conclusion, a new property of CNTs has been observed for the first time during the synthesis of semiconductor NPs. CNTs trigger a morphological transformation of CdSe nanorods into pyramidal-shaped nanoparticles and a tight attachment to them. The CNT-NP interaction can be understood as a ligand exchange where nanotubes stabilize the NPs. The released ligands (ODPA) promote the transformation of the NPs by selective etching and Ostwald ripening mechanism. The presented non-covalent attachment between CNTs and NPs should, furthermore, be most advantageous in order to combine the outstanding electrical properties of CNTs with the unique possibility of bandgap tuning of quantum dots. We, therefore, expect a large impact of these materials in optoelectronics and photovoltaics. Beside this, we also expect a significant improvement of the mechanical properties in CNT- polymer composites. Due to the roughening of the surface of the nanotubes by the attached NPs the interface with the surrounding polymer matrix should be significantly improved and the contact to the polymer should be strengthened without destroying the extraordinary mechanical properties of the nanotubes by surface functionalization.

\section*{Acknowledgement}

This work is partially supported by the European Commission through a Marie Curie Intraeuropean Fellowship. The authors thank Melanie Rappold for experimental help.

\clearpage

\end{document}